\documentclass[aps,nofootinbib,twocolumn,superscriptaddress,floatfix]{revtex4}

\usepackage{color} %

\usepackage{amsmath,amssymb} %
\usepackage{graphicx} %
\usepackage{times} %

\IfFileExists{srcltx.sty}{\usepackage[active]{srcltx}}

\usepackage{bm,paralist,xspace,array}

\renewcommand{\S}{\mathcal{S}} 
\renewcommand{\lim}{\mathrm{lim}} %
\newcommand{\mpc}{\:\mathrm{Mpc}} 
\newcommand{\ta}{\mathrm{ta}} %
\newcommand{\parfrac}[2]{\left(\frac{#1}{#2}\right)} %
\usepackage{hyperref}

\IfFileExists{hypernat.sty}{\usepackage{hypernat}}

\begin{document}

\title{New probe of modified gravity}

\author{A.~Boyarsky}%
\affiliation{Ecole Polytechnique F\'ed\'erale de Lausanne, FSB/ITP/LPPC, BSP
  720, CH-1015, Lausanne, Switzerland} \affiliation{Bogolyubov Institute of
  Theoretical Physics, Kyiv, Ukraine}

\author{O.~Ruchayskiy}%
\affiliation{Ecole Polytechnique F\'ed\'erale de Lausanne, FSB/ITP/LPPC, BSP
  720, CH-1015, Lausanne, Switzerland} \date{January 4, 2010}

\begin{abstract}
  We suggest a new efficient way to constrain a certain class of large scale
  modifications of gravity. We show that the scale-free relation between
  density and size of Dark Matter halos, predicted within the $\Lambda$CDM
  model with Newtonian gravity, gets modified in a wide class of theories of
  modified gravity.
\end{abstract}

\maketitle

Models with the large scale modification of gravity are actively discussed in
the recent years in connection with the observed accelerated expansions of the
Universe and because they can be related to the existence of extra
dimensions~\cite{DGP,Deffayet:01a}. However, the general principles of gauge
invariance and unitarity strongly constrain possible theories of gravity,
modifying the Newton's law at large distances (see analysis
of~\cite{Dvali:06a}). Thus, in addition to its phenomenological applications,
this problem is related to the fundamental questions of particle physics,
field theory and gravity.  It is therefore important to search for large scale
modifications of gravity experimentally.

A possible set of consistent (as a spin-2 field theory) large scale
modifications of gravity is described by two parameters -- scale $r_c$ and a
number $0\le \alpha < 1$~\cite{Dvali:06a,Dvali:07a,Dvali:10a}.  $r_c$ marks
the distances at which at the \emph{linearized level} gravitational law
changes from $1/r^2$ to some other power $1/r^n$, and the parameter $\alpha$
determines the value of $n$.  Phenomenologically, deviations from Newton's law
we are looking for may be represented in this parameter space. A significant
fraction of this space is excluded by precision measurements of the Moon
orbit~\cite{Dvali:02b}. Other natural probes of such modifications are
cosmological observables (see
e.g.~\cite{Dvali:03a,SDSS-II,Lombriser:09a,Scoccimarro:09a,Chan:09a,Khoury:09a}
and refs. therein).

In this work we identify a new observable sensitive to the large scale
modifications of gravity. We demonstrate that universal properties of
individual dark matter halos are also affected by the modifications of
gravity, and provide novel way to probe them. Namely, we show that the
scale-invariant relation between density and size of dark matter halos,
predicted by Newtonian gravity within the $\Lambda$CDM
model~\cite{Boyarsky:09c} and found to hold to a good precision in observed
dark matter halos~\cite{Boyarsky:09b}, may receive non-universal
(size-dependent) corrections for a wide range of parameters $r_c$ and
$\alpha$.

Formation of structures in the Universe is an interplay between gravitational
(Jeans) instability and overall Freedman expansion. The gravitational collapse
does not start until the potential energy $\mathcal{U}$ of a gravitating dark
matter system overpowers the kinetic energy of the Hubble expansion
$\mathcal{K}\sim \frac 12 H^2 R^2$.  Once the gravitational collapse has
began, at any moment of time $t$ a dark halo is confined within a sphere of
zero velocity or a \emph{turn-around sphere}. As Hubble expansion rate $H(t)$
decreases with time, the turn-around radius $R_\ta(t)$ grows. In the Newtonian
cosmology with potential $\phi_N(r) = - GM/r$ the turn-around radius $R_\ta$
is
\begin{equation}
  \label{eq:1}
  R_\ta \propto \parfrac{GM}{H^2}^{1/3}
\end{equation}
(today for masses $\sim 10^{12} M_\odot$ the turn-around radius is $\sim
1\mpc$).  Notice that at any moment of time the average density within a
turn-around radius~(\ref{eq:1}) is proportional to the cosmological density
and is the same for halos of all masses:
\begin{equation}
  \label{eq:2}
  \rho_\ta \propto \frac{H^2}{G} \propto \bar\rho_{tot}(t)
\end{equation}
It was shown in~\cite{Boyarsky:09c} that the property~(\ref{eq:2}) leads to a
universal relation between characteristic scales and densities of dark matter
halos. This relation holds in wide class of dark-matter dominated objects
(from dwarf galaxies to galaxy clusters)~\cite{Boyarsky:09b} (see
also~\cite{Kormendy:04,Donato:09,Gentile:09a}). The relation is in a very good
agreement with pure dark matter simulations~\cite{Maccio:08,Springel:08a},
suggesting that baryonic feedback can be neglected in this case. Therefore,
this relation can serve as a new tool of probing properties of dark matter and
gravity at large scales.

The relation~(\ref{eq:2}) continues to hold in the Universe where gravity is
modified by the cosmological constant $\Lambda$. The gravitational energy of a
body of mass $M$ at distance $r$ becomes $\mathcal{U}_\Lambda = -\frac{GM}r
-\frac{\Lambda r^2}6$. Comparing it with the kinetic energy of the Hubble flow
$\mathcal{K}$ one arrives once again to the relation~(\ref{eq:2})
\begin{equation}
  \label{eq:3}
  \rho_\ta(t) \propto \frac{\Lambda}{G}
\end{equation}
(c.f.~\cite{Boyarsky:09c}). The relation~(\ref{eq:1}) still holds and is again
independent on the mass of the halo.

What is the most general form of gravitational potential, for which the
property~(\ref{eq:2}) remains true? Clearly, it will hold for all the
gravitational potentials of the form
\begin{equation}
  \label{eq:4}
  \phi(r) = -\frac{GM}{r}F\parfrac{\rho(r)}{\rho_\star}
\end{equation}
where $\rho_\star$ is some constant with dimension of density.  In particular,
$\Lambda$-term obeys this property (with $\rho_\star \propto \Lambda/G$). All
the theories of the form~(\ref{eq:4}) obey the property that relative
correction to the Newtonian potential $\phi_N$ depends \emph{only} on the
density $\rho(r)$ within a radius $r$ (and not on the mass or the size of
objects).

Next, we consider the modifications of gravity~\cite{Dvali:06a,Dvali:07a}.
The gravitational potential of a spherically symmetric system of mass $M$
there has the form
\begin{equation}
  \label{eq:5}
  \phi_\alpha(r) = -\frac{GM}r \pi(\frac r{r_V})
\end{equation}
where $0\le \alpha < 1$. Here the characteristic (\emph{Vainstein}) radius
$r_V$ is defined as~\cite{Deffayet:2001uk,Dvali:06a,Dvali:07a}
\begin{equation}
  \label{eq:6}
  r_V = \Bigl( 2G M r_c^\beta\Bigr)^\frac1{1+\beta}\quad\text{where }\beta = 4(1-\alpha)
\end{equation}
If the scale $r_c$ is of the order of $ \sim H_0^{-1}$, such modifications of
gravity can provide an explanation for the late-time cosmological expansion of
the Universe~\cite{DGP,Deffayet:01a}.  The corrections to Newton's law become
negligible as $r\to0$ ($\pi(0)=1$) and the radius~(\ref{eq:6}) characterizes
the scale where the deviations from Newton's potential become of order unity.
Using the relation
\begin{equation}
  \label{eq:7}
  \frac{r}{r_V} = \left(\frac{r^{1+\beta}}{2 G M
      r_c^{\beta}}\right)^\frac1{1+\beta} 
  \propto M^{\frac{\beta-2}{3(1+\beta)}}\frac{1}{\rho^{1/3}\bigl(2 G r_c^\beta \bigr)^{\frac1{1+\beta}}}
\end{equation}
(where $\rho = M/r^3$) we find that among the theories of modified
gravity~(\ref{eq:5}) only $\beta=2$ ($\alpha=\frac12$, the DGP
model~\cite{DGP}) possess the property~(\ref{eq:4}) and
consequently~(\ref{eq:2}).

\begin{figure}
  \centering
  \includegraphics[width=\linewidth]{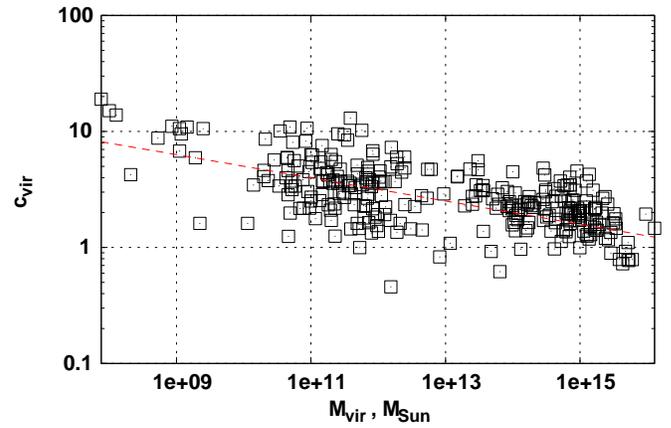}
  \caption{Comparison of $c_{\rm vir}=R_{\rm vir}/r_{\rm C}$ as a function of
    DM halo mass for observed DM density profiles from the catalog
    of~\cite{Boyarsky:09b}.}
  \label{fig:c-m}
\end{figure}

The property~(\ref{eq:2}) can be probed experimentally.  Extensive catalog of
DM-dominated objects of all scales, collected in~\cite{Boyarsky:09b}, exhibits
a simple scaling relation of the properties of the DM halos.  Dark matter
distribution in the majority of observed objects can be described by one of
the universal DM profiles (such as e.g. NFW~\cite{Navarro:96} or
Burkert~\cite{Burkert:95}). Such profiles may be parametrized by two numbers,
directly related to observations -- a characteristic radius $r_{\rm C}$
(off-center distance where the rotation curve becomes approximately flat,
equal e.g. to $r_s$ for NFW) and a DM central mass density $\bar \rho_{\rm
  C}$, averaged inside a ball with the size $r_{\rm C}$.  It was shown that DM
column density $\S \propto \bar\rho_{\rm C} r_{\rm C}$,
(see~\cite{Boyarsky:09b,Boyarsky:09c} for a detailed definition) changes with
the mass as $\S \propto M^\kappa$, where $\kappa\approx 0.22-0.33$.  It was
demonstrated in~\cite{Boyarsky:09c} that in the simplest self-similar model
(i.e. assuming that $r_{\rm C}/R_\ta$ is the same for the DM halos of all
masses) property~(\ref{eq:1}--\ref{eq:2}) implies a scaling $\S \propto
M^{1/3}$, compatible with observations.  Observations (see
e.g.~\cite{Vikhlinin:05,Buote:06,Schmidt:07,Comerford:07,Mandelbaum:08})
demonstrate that the ratio of $r_{\rm C}$ to the virial radius depends weakly
(as $M^{\approx -0.1}$) on the mass of DM halos. The Fig.~\ref{fig:c-m} shows
ratio of the virial radius of a halo to its $r_{\rm C}$ for DM density
profiles from the catalog of~\cite{Boyarsky:09b}.  These results are in
perfect agreement with the $\Lambda$CDM numerical simulations~ (see
e.g.~\cite{Neto:07,Maccio:08}).  Due to this slight deviation from the
self-similarity, the best fit value of the scaling parameter $\kappa=0.23$
(see~\cite{Boyarsky:09c} for discussion).  For the qualitative discussion of
this work, it is important that $\S(M)$ is a featureless power-law dependence,
whose slope does not depend on mass and that the deviation from the slope
$\frac13$ is small (as follows from observations).

We can conclude that in the theories, satisfying the condition~(\ref{eq:5})
(e.g. in DGP model) the properties of DM halos theory will follow the same
scaling relation $\S \propto M^{1/3}$ and the difference with the $\Lambda$CDM
case will only be in a different normalization of this scaling relation.  For
other theories, described in~\cite{Dvali:06a,Dvali:07a}, with $\alpha \neq
\frac12$, we can see from the Eq.~(\ref{eq:7})) that the potential
$\phi_\alpha(r)$ is not of the form~(\ref{eq:4}) and we can expect deviation
from the universal scaling law.

\section{$\S-M$ relation for general $\alpha$}
\label{sec:scaling-alpha-neq}

Let us work out the $\S-M$ for a general $\alpha$ in details.  A general
expression, relating the turn-around time, turn-around radius and mass within
this radius follows from energy conservation and is given by
\begin{equation}
  \label{eq:8}
  t_0 = \frac1{\sqrt 2}\int^{R_\ta}_0 \frac{dr}{\sqrt{\phi_\alpha(r)-\phi_\alpha(R_\ta)}}
\end{equation}
Using the general form~(\ref{eq:5}) we can rewrite the expression~(\ref{eq:8})
in the following form:
\begin{equation}
  \label{eq:9}
  t_0 =  \parfrac{\pi^2 R_\ta^3}{8 G M}^{1/2} I(x_\ta)
\end{equation}
where dimensionless ratio $x_\ta \equiv R_\ta/r_V$ and the function $I(x_\ta)$
is given by
\begin{equation}
  \label{eq:10}
  I(x_\ta) = \frac2\pi \int^1_0 \frac{dx}{\left(\frac{\pi(x x_\ta)}x-\pi(x_\ta)\right)^{1/2}}
\end{equation}
The solution of this equation gives us the ``density'' $\rho_\ta \equiv
M/R_\ta^3$ as a function of $M$. When $r_c \to \infty$ the turn-around density
$\rho_\ta$ becomes
\begin{equation}
  \label{eq:11}
  \rho_\ta = \frac{M}{R^3_\ta} = \frac{\pi^2}{8 G t_0^2} \equiv \rho_0
\end{equation}
Here the \emph{constant} $\rho_0$ is a function of lifetime of the Universe
only and does not depend on parameters of a dark matter halo.  This gives a
desired relation between a turn-around density and the life-time of the
Universe in the pure Newtonian cosmology (without cosmological constant).  The
function $I(x)$ is defined in such a way that in the Newtonian limit
$\pi(x)=1$ one gets $I(x)=1$.

The derivation of Eq.~(\ref{eq:9}) demonstrates that for all theories of
gravity of the form~(\ref{eq:4}) (including $\Lambda$-term and the DGP model)
$\rho_\ta$ is a function of $\rho_0$ \emph{only} and does not depend on the
mass/size of a particular halo. As a result $\S\propto M^{1/3}$
(see~\cite{Boyarsky:09c} for details). 

Further analysis of Eq.~(\ref{eq:9}) depends on the form of the function
$\pi(x)$. Its exact form is not known (apart from the DGP case). Let us start
with analyzing several limiting cases.

If the Vainstein radius $r_V \ll R_\ta$ for halos of all masses that are
experimentally observed (roughly from $\sim 10^8 M_\odot$ to $\sim
10^{16}M_\odot$), then for distances $r\gg r_V$ the corrections to the
Newtonian potential reduce either to the order one renormalization of the
gravitational constant (on the ``normal branch'') or become indistinguishable
from the $\Lambda$-term (``self-accelerated branch''). In both cases $\S(M)
\propto M^{1/3}$ with the normalization, different from the pure Newtonian
case.

\begin{figure}[t]
  \centering
  \includegraphics[width=.4\textwidth]{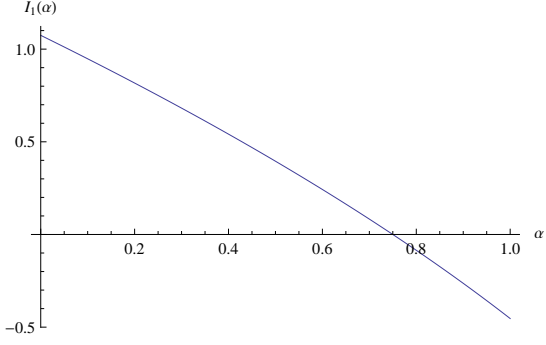}
  \caption{Function $I_1(\alpha)$ defined in Eq.~(\protect\ref{eq:13})}
  \label{fig:i1}
\end{figure}

In the opposite case $r_V \gg R_\ta$, one can utilize the perturbative
expansion of the function $\pi(x)$. The gravitational potential well inside
the Vainstein radius is given by~\cite{Dvali:06a,Dvali:07a}
\begin{equation}
  \label{eq:12}
  \pi(x\ll 1) \approx 1 + c_1 x^a\quad ;\quad
  a = \frac{\beta+1}2=\frac{5-4\alpha}2
\end{equation}
where $c_1\sim \mathcal{O}(1)$ and is positive for the ``normal branch'' and
negative for the ``self-accelerating branch''~\cite{Dvali:10a} in full analogy
with the DGP model. Notice that $a>1$ for $\alpha < \frac34$. Using
expansion~(\ref{eq:12}) one arrives to
\begin{equation}
  \label{eq:13}
  I(x_\ta\ll 1) \approx 1 + \frac{c_1}2 x_\ta^a\underbrace{\left[\frac2\pi\int^1_0
      dx\,\frac{1 - x^{a-1}}{\Bigl(\frac1x-1\Bigr)^{3/2}}\right]}_{\equiv I_1(\alpha)}
\end{equation}
where the function $I_1(\alpha)$ is shown on the Fig.~\ref{fig:i1}.

Substituting the expression~(\ref{eq:13}) back into equation~(\ref{eq:9}) and
using~(\ref{eq:7}), we obtain
\begin{equation}
  \label{eq:14}
  \parfrac{\rho_0}{\rho_\ta}^{1/2}\Bigl(1+\frac{c_1}2 x_\ta^a I_1(\alpha)\Bigr)=1
\end{equation}
As $x_\ta \ll 1$ and $a>1$, one finds that
\begin{equation}
  \label{eq:15}
  \rho_\ta \simeq \rho_0 \Bigl(1+c_1 I_1(\alpha) x_\ta^a(\rho_0)\Bigr)
\end{equation}
where to compute $x_\ta$ we use Eq.~(\ref{eq:7}) with $ \rho_0$ instead of
$\rho_\ta$. From Eqs.~(\ref{eq:7}) and~(\ref{eq:15}) we see once again that
for all $\alpha \neq \frac12$ (i.e. $\beta \neq 2$) the turn-around density
$\rho_\ta$ loses its universality and becomes the function of the halo mass
$M$. The turn-around radius $R_\ta(M)$ is related to $\rho_\ta$ via $R_\ta(M)
= (M/\rho_\ta)^{1/3}$.

\begin{figure}
  \centering
  \includegraphics[width=\linewidth]{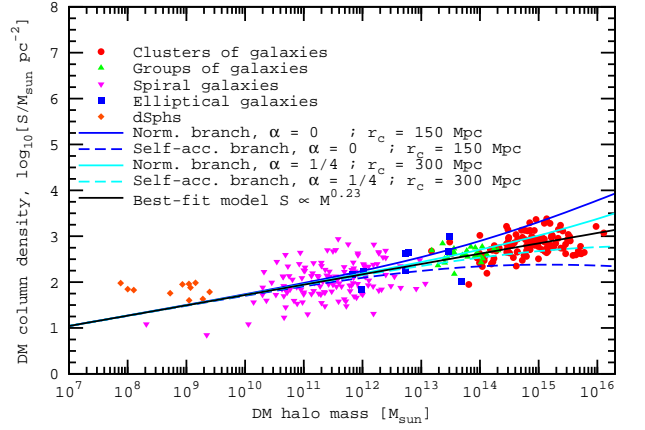}
  \caption{Examples of $\S-M$ relation in theories with $\alpha < 1/2$
    together with the data from~\protect\cite{Boyarsky:09b}. }
  \label{fig:S-M1}
\end{figure}

Under the assumption of exact self-similarity, discussed above (i.e. $r_{\rm
  C}/R_\ta = \mathrm{const}$) one arrives to the following expression for $\S$
(recall that $\beta = 4(1-\alpha)$):
\begin{align}
  \label{eq:17}
  \S(M) & = \rho_{\rm C} r_{\rm C} \propto M^{1/3} \rho^{2/3}_\ta \\
  &\propto M^{1/3}\rho_0^{2/3} \left(1+ \frac 23 c_1 I_1(\alpha)
    \parfrac{M}{M_{\lim}}^{\frac{1-2\alpha}{3}}\right)\label{eq:18}
\end{align}
where 
\begin{align}
  \label{eq:16}
  M_{\lim} \equiv
  \frac1G\left[\parfrac{r_c}{2}^{3\beta}\parfrac{\pi}{t_0}^{2(1+\beta)}\right]^{\frac1{\beta-2}}
\end{align}
An example of the relation~(\ref{eq:18}) for several $\alpha$'s and $r_c$ is
shown in Fig.~\ref{fig:S-M1}.

\begin{figure}
  \centering
  \includegraphics[width=\linewidth]{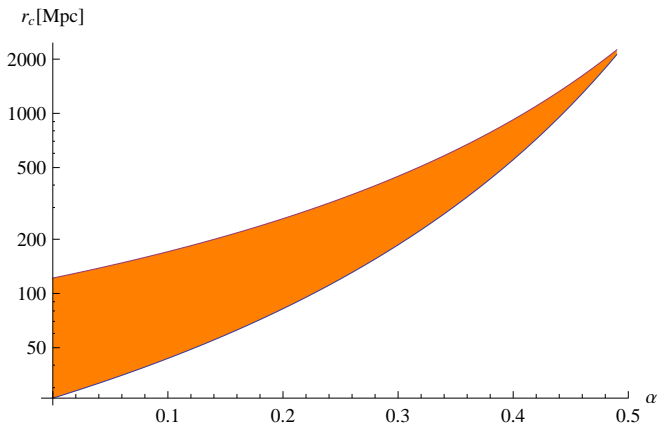}
  \caption{Range of $r_c$ for which $r_V\approx R_\ta$ for some halo masses in
    the range $10^{10}-10^{15} M_\odot$.}
  \label{fig:rclim}
\end{figure}

Clearly, the most interesting case is when $r_V \approx R_\ta$ for some range
of observed halo masses. In this regime the deviations from Newtonian gravity
become the strongest.  The range of values $r_c$ for which this happens is
shown in Fig.~\ref{fig:rclim} (the value of $t_0$ is chosen to be the lifetime
of the Universe $t_0 \simeq 1.3\times 10^{10}$~years). We expect that for
$r_c$ in the region Fig.\ref{fig:rclim} the slope of the relation $\S\propto
M^\kappa$ will change.  Analysis of this case requires however an exact
solution of the non-linear analog of the Poisson equation in theories with
$\alpha$ (see e.g.~\cite{Dvali:07a}), i.e. the knowledge of properties of the
function $\pi(r/r_V)$ in the range of radii, where the perturbative
expansion~(\ref{eq:12}) breaks down. Notice, that the region where $r_V
\approx R_\ta$ shrinks toward the value $r_c = \frac 2\pi t_0$ as $\alpha \to
\frac12$. For this value of $r_c$ the Vainstein radius in the DGP model is
equal to the turn-around radius \emph{for all halo masses}.\footnote{In
  general for $r_c \sim H_0^{-1}$ the turn-around radius is smaller than $r_V$
  for $\alpha < \frac 12$ and bigger than $r_V$ for $\alpha > \frac 12$.}
Knowledge of the potential $\pi(x)$ would also allow to probe the
modifications of gravity directly in the Local Group and other nearby
galaxies, by studying the infall trajectory around the turn-around radius (see
e.g.~\cite{Steigman:98}).

Another way to probe the $\S-M$ relation for general $r_c$ and $\alpha$ is to
do numerical simulations in the theories of modified gravity (see examples
in~\cite{Chan:09a,Khoury:09a}) and compare directly with observations both the
scaling of the central values of $\S$ and $M$ and the scatter around it.

\textbf{Conclusion.}  The main purpose of this work was to identify a new
observable that can be used to constrain the large scale modifications of
gravity.  We see that the scaling properties of dark matter halos are
sensitive to such modifications.  We demonstrated that the models with $\alpha
\neq \frac12$ predict the deviation from a simple power-law scaling in the
$\S(M)$ relation.  Comparison of predictions of such models with the data,
collected in~\cite{Boyarsky:09b} potentially allows to restrict the values of
$r_c$ from below for a given $\alpha$.  The improved data-processing and new
observational data on DM distributions will allow to strengthen these bounds
and make them quantitative.

In this work we have analyzed only the case when the turn-around sphere is
well inside the Vainstein radius, $r_V \gg R_\ta$.  To analyze a general case,
a better theoretical understanding of the function $\pi(r)$ (defined via
Eq.~(\ref{eq:5})) is needed. Together with better quality of data this will
allow to extend our analysis to a wider range of parameters.

In the case $\alpha = \frac12$ (the DGP model) the $\S(M)$ dependence remains
featureless. In this case one has $r_V \approx R_\ta$ for all masses (for $r_c
\sim H_0^{-1}$) and the deviations from the Newtonian gravity at turn-around
radius will be strong.  Therefore this model (in general, all models that have
$r_V \sim R_\ta$ for halos of $M\sim 10^{12} M_\odot$) can be probed by
studying the infall trajectories around the turn-around radius in the Local
Group and nearby galaxies~\cite{Steigman:98} using the available data and the
data from forthcoming surveys of the Milky Way as well as GAIA mission.

\textbf{Acknowledgments.} We would like to thank G.~Dvali for many useful and
illuminating discussions and careful reading of the manuscript. This work was
supported in part by the Swiss National Science Foundation.

\let\jnlstyle=\rm\def\jref#1{{\jnlstyle#1}}\def\aj{\jref{AJ}}
\def\araa{\jref{ARA\&A}} \def\apj{\jref{ApJ}\ } \def\apjl{\jref{ApJ}\ }
\def\apjs{\jref{ApJS}} \def\ao{\jref{Appl.~Opt.}} \def\apss{\jref{Ap\&SS}}
\def\aap{\jref{A\&A}} \def\aapr{\jref{A\&A~Rev.}} \def\aaps{\jref{A\&AS}}
\def\azh{\jref{AZh}} \def\baas{\jref{BAAS}} \def\jrasc{\jref{JRASC}}
\def\memras{\jref{MmRAS}} \def\mnras{\jref{MNRAS}\ }
\def\pra{\jref{Phys.~Rev.~A}\ } \def\prb{\jref{Phys.~Rev.~B}\ }
\def\prc{\jref{Phys.~Rev.~C}\ } \def\prd{\jref{Phys.~Rev.~D}\ }
\def\pre{\jref{Phys.~Rev.~E}} \def\prl{\jref{Phys.~Rev.~Lett.}}
\def\pasp{\jref{PASP}} \def\pasj{\jref{PASJ}} \def\qjras{\jref{QJRAS}}
\def\skytel{\jref{S\&T}} \def\solphys{\jref{Sol.~Phys.}}
\def\sovast{\jref{Soviet~Ast.}} \def\ssr{\jref{Space~Sci.~Rev.}}
\def\zap{\jref{ZAp}} \def\nat{\jref{Nature}\ } \def\iaucirc{\jref{IAU~Circ.}}
\def\aplett{\jref{Astrophys.~Lett.}}
\def\apspr{\jref{Astrophys.~Space~Phys.~Res.}}
\def\bain{\jref{Bull.~Astron.~Inst.~Netherlands}}
\def\fcp{\jref{Fund.~Cosmic~Phys.}} \def\gca{\jref{Geochim.~Cosmochim.~Acta}}
\def\grl{\jref{Geophys.~Res.~Lett.}} \def\jcp{\jref{J.~Chem.~Phys.}}
\def\jgr{\jref{J.~Geophys.~Res.}}
\def\jqsrt{\jref{J.~Quant.~Spec.~Radiat.~Transf.}}
\def\memsai{\jref{Mem.~Soc.~Astron.~Italiana}}
\def\nphysa{\jref{Nucl.~Phys.~A}} \def\physrep{\jref{Phys.~Rep.}}
\def\physscr{\jref{Phys.~Scr}} \def\planss{\jref{Planet.~Space~Sci.}}
\def\procspie{\jref{Proc.~SPIE}} \let\astap=\aap \let\apjlett=\apjl
\let\apjsupp=\apjs \let\applopt=\ao

\end{document}